\documentstyle[11pt]{article}
\input epsf

\newcommand{\eps}{{\epsilon}}
\newcommand{\om}{{\omega}}
\newcommand{\Om}{{\Omega}}

\newcommand{\pa}{{\partial}}
\newcommand{\beq}{\begin{equation}}
\newcommand{\eeq}{\end{equation}}
\newcommand{\beqa}{\begin{eqnarray}}
\newcommand{\eeqa}{\end{eqnarray}}
\newcommand{\bi}{\begin{itemize}}
\newcommand{\ei}{\end{itemize}}
\newcommand{\vr}{{\vec r}}

\newcommand{\lsim}{\hbox{ {\raisebox{0.06cm}{$<$} \raisebox{-0.14cm}{$\!\!\!\!\!\!\!\!\: \sim$}} } }
\newcommand{\rsim}{\hbox{ {\raisebox{0.06cm}{$>$} \raisebox{-0.14cm}{$\!\!\!\!\!\!\!\!\: \sim$}} } }
 \title{Paramagnetic-diamagnetic interplay in quantum dots for non-zero
temperatures}

\author{ Yu.~P.~Krasny$^1${\thanks{Corresponding author,
 e-mail krasnyj@math.uni.opole.pl}},
N.P.~Kovalenko$^2$, U.~Krey$^3$, L.~Jacak$^4$\\ 
 \\ $^1$
   Department of Mathematics, University of Opole, 45-052 Opole, Poland 
 \\ $^2$ Department of Physics, Odessa State University, 270026 Odessa,
Ukraine
\\ $^3$ Institute of Physics II, University of Regensburg, 93040 Regensburg,
Germany
\\ $^4$ Institute of Physics, Wroclaw University of Technology, 50-370
Wroclaw, Poland
  }

\date{received Nov.\ 13, 2000, by J Phys CM; revised March 25, 2001;
 accepted March 29,
2001}

\input epsf
\begin{document}

\large
\maketitle
\begin{abstract}

\noindent In the usual Fock-and Darwin-formalism with parabolic potential
characterized by the confining energy $\eps_0=\hbar\om_0\approx 3.4$ meV,
but including explicitly also the Zeeman coupling between spin and magnetic
field, we study the combined orbital and spin magnetic properties of quantum
dots in a two-dimensional electron gas with parameters for GaAs, for $N$ =1
and $N\gg 1$ electrons on the dot.\newline
 For $N=1$ the magnetization $M(T,B)$ consists of a paramagnetic spin
contribution and a diamagnetic orbital contribution, which dominate in a
non-trivial way at low temperature and fields rsp.\ high
temperature and fields.\newline
For $N\gg 1$, where orbital and spin effects are intrinsically
 coupled in a subtle way  and cannot be separated, we find in a simplified
 Hartree approximation
that at $N=m^2$, i.e.\ at a half-filled last shell, $M(T,B,N)$
is parallel (antiparallel) to the magnetic field,
 if temperatures and fields are low enough (high enough), whereas for
$N\ne m^2$ the magnetization oscillates  with $B$ and $N$ as a $T$-dependent
periodic function of the variable $x:=\frac{\sqrt{N}eB}{2m^*c\cdot\om_0}$, with
$T$-independent  period $\Delta x =1$ (where $m^*=0.067 \,m_0$ is the small
effective mass of GaAs, while $m_0$ is the electron mass).\newline
Correspondingly, by an {\it adiabatic demagnetization process},
 which should only be fast enough with respect to the slow 
{\it transient time} of the magnetic
properties  of the dot, the temperature of the dot diminishes
 rsp.\ increases with decreasing magnetic field, and in some cases 
we obtain quite pronounced effects.
\end{abstract}
\vspace*{0.5cm}

\section{Introduction}

Besides the {\it charge} degrees of freedom, the {\it spin} of the
electrons in quantum dots will certainly play an important role
in future magneto-electronic devices for classical or quantum computing,
   involving quantum dots
 ('artificial atoms'), \cite{refLoss}, although 
the spin degrees of freedom are usually neglected,
 since typically the {\it orbital} magnetism dominates in quantum dots,
 as is known, and
 as we also will see below. However, in this  paper
 we look at the {\it magnetic} properties of quantum dots more in detail,
 including the 'atypical' spin degrees of freedom, to see
whether in this way one  may be lead to  some 'new physics'.
 Moreover,
it is clear that for our purpose {\it not} 
the most elaborate many-body techniques are important, but
simple approaches should suffice to draw  relevant conclusions.
 With this in mind, we concentrate below 
on the two case $N=1$ and $N\gg 1$, where $N$ is the number of
electrons in the dot. 

In any case, solids with quantum dots
 (i.e.\ planar artificial atoms) being placed in an
external magnetic field $\vec B$ have to acquire an additional magnetic
moment. If the dots do not interact, this moment is $N_D\cdot\vec M$, where
$\vec M$ is defined as 
the mean magnetic moment of a {single} dot and $N_D$ the number of dots.
That is why the following calculations reduce to considering the behaviour
of a {\it single} dot being in thermodynamic equilibrium with the
surrounding. In this case we can consider the magnetic field $\vec B$ acting
on the electrons in the dot as being identical to the external field.

In the  following we always assume that the field $\vec B$ is constant in space
and has the $z$-direction, whereas the electrons move in the $(x,y)$-plane.
\section{The case of $N=1$:}
For the beginning, we consider the simple case of a dot with one electron.
In such a model, and in the usual effective-mass approximation, the motion
of the electron is described by the Hamiltonian
\beq\label{eq1}
\hat{\cal H}= -\frac{\hbar^2}{2m^*}\nabla^2 +\frac{1}{2}m^*\omega^2 r^2
+\frac{\hbar\omega_c}{2}\left (\hat l_z+g^*\hat S_z \right)\,,
\eeq
where $m^*$ is the effective mass of the electron ($=0.067\cdot m_0$ for GaAs,
where $m_0$ is the electronic mass), $\nabla^2$ is the Laplacian in two
dimensions, $\hbar=h/(2\pi)$, with Planck's constant $h$,
$\omega_c=\frac{|e| B}{m^*c}$ is the cyclotron fequency,
$\omega^2:=\omega_0^2 + \frac{\omega_c^2}{4}$, where $\omega_0$ is the
parameter characterizing the strength of the parabolic potential, which
essentially  confines  the electron to the dot. Finally, $\hat
l_z=-i(x\frac{\partial}{\partial y} -y\frac{\partial}{\partial x})$, with
integer eigenvalues $m$, is the (reduced) operator of the {\it angular momentum}, while the
corresponding (reduced) {\it spin-momentum} operator $\hat S_z$ has the eigenvalues
$\pm \frac{1}{2}$ ( Here 'reduced' means 'measured in units of $\hbar$').
 Furthermore, in the following we use the 'effective Bohr
magneton' $\mu_B^* :=\frac{\hbar \omega_c}{2B}=\frac{\hbar|e|}{2\,m^*c}$\,;
 $g^*:=\frac{m^*}
{m_0}\cdot g$ is
the corresponding effective $g$-factor, where for the free electron
one would have $m^*=m_0$ and $g=2$, whereas for GaAs, we have $g\cong
-0.44$, and (as already mentioned) $m^*\cong 0.067\, m_0$. 
Therefore, for GaAs, the quantity $|g^*| $ is $\ll 1$, whereas $\mu_B^*\gg
\mu_B$.

\noindent The eigenfunctions of the Hamiltonian (\ref{eq1}) have the following form,
\cite{refJacak}
\beq\label{eq2}
\psi(\vec r,\sigma )=\varphi_{n_+,n_-}(\vec r)\chi_{s_z}(\sigma)\,,
\eeq
where $\chi_{s_z}(\sigma )$ are normalized eigenfunctions of the spin
operator $\hat S_z$ with eigenvalue $s_z=\pm \frac{1}{2}$, while the
coordinate wave function can be written as
\beq\label{eq3}
\varphi_{n_+,n_-}(\vec r)=\frac {(\hat a)^{n_+}\cdot (\hat b)^{n_-}}
{l_0 \sqrt{2\pi\,n_+!\,n_-!}}\,\exp\left (-\frac{-|x+{\rm
i}y|^2}{2l_0^2}\right )\,. 
\eeq
Here
\beqa
\hat a &:=& \frac{1}{2{\rm i}}\left [\frac{x+{\rm i}y}{l_0}
-l_0(\frac{\partial}{\partial x}+{\rm i}\frac{\partial}{\partial y}) \right ]\nonumber\cr
\hat b &:=& \frac{1}{2} \,\,\left [ \frac{x-{\rm i}y}{l_0} 
-l_0(\frac{\partial}{\partial x}-{\rm i}\frac{\partial}{\partial y})\right
]\,,\nonumber
\eeqa
with $l_0^2=\frac{\hbar^2}{m^*\omega}$;\, $n_{\pm}=0,1,2,...$;\,
$s_z=\pm\frac{1}{2}$. 

The energy eigenvalues corresponding to these wavefunctions are
\beq \label{eq4}
\epsilon_{n_+,n_-,s_z}=\epsilon_+\cdot (n_++\frac{1}{2})
+\,\epsilon_-\cdot (n_-+\frac{1}{2})+g^*\cdot\frac{\hbar\omega_c}{2}\cdot s_z\,,
\eeq
with \quad $\epsilon_{\pm}:=\hbar\omega\,\pm\,\frac{1}{2}\hbar\omega_c$.

The partition function $Z(T,B)$ of the system can be easily calculated and
is equal to
\beqa \label{eq5}
Z(T,B)&=&\sum\limits_{n_+=0}^\infty\sum\limits_{n_-=0}^\infty
\sum\limits_{s_z=-1/2}^{1/2}\,\exp(-\frac{\epsilon_{n_+,n_-,s_z}}{k_BT})
\cr
&=& {\rm ch}(\frac{g^*\hbar\omega_c}{4k_BT})\cdot \left [
{\rm ch}(\frac{\hbar\omega}{k_BT})-{\rm ch}(\frac{\hbar\omega_c}{2k_BT})
\right ]^{-1}\,,
\eeqa
where $k_B$ is the Boltzmann constant and $T$ the 'Kelvin temperature'. All
other characteristic thermodynamic quantities can be found from $Z(T,B)$ by
known derivatives. For instance, the mean moment of a dot is,
\cite{refVonsovskij}:
\beq \label{eq6} 
M=k_BT\frac{1}{Z}\left (\frac{\partial Z}{\partial B}\right )_T\,.\eeq

Thus for $N=1$, the Free Enthalpy\,\,\,\, $G(T,B)=-k_BT\cdot \ln Z(T,B)$, and the
magnetization as well, can simply be separated into a 'para\-magnetic'
 spin contribution, corresponding to the first factor on the r.h.s.\ of
(\ref{eq5}), and the usual diamagnetic 'orbital' contribution corresponding
to the second factor in (\ref{eq5}). Due to the smallness of $g^*$, the
para\-magnetic contribution is very small in GaAs.
However at low fields, for high enough temperatures  the spin contribution
dominates in any case, since a systematic Taylor expansion shows that the
'paramagnetic factor' is
 $Z_{\rm spin}\cong 1+\frac{(g^*)^2e^2B^2}{8(m^*)^2k_B^2T^2}+...$ (i.e.\ the
correction is $\propto B^2/T^2$), whereas
the 'orbital factor' is  $Z_{\rm orbital}\cong 1
-\frac{\hbar^4\om_0^2\om_c^2}{48\, k_B^4T^4}+...$ (i.e.\ here the correction
is $\propto B^2/T^{4}$). Note that here we have explicitly used the
$B$-dependence of $\om=\sqrt{\om_0^2+\frac{\om_c^2}{4}}$, which sometimes should not be
'approximated away' too early.

Of course we are more interested in the low-temperature behaviour: In any
case, the magnetization $M(T,B)$ can be calculated for $N=1$ completely generally
from the following formula, with $\beta:=(k_BT)^{-1}$, and
with the characteristic energy $\eps_B\propto B$, namely
$\eps_B:=\frac{\hbar\,\om_c}{2}$:
\beq
\label{eqM(T,B)}
\frac{M(T,B)}{\mu_B^*}=\frac{g^*}{2}\cdot{\rm th}\left (
\frac{g^*\eps_B\,\beta}{2}\right )
- \frac
{{\rm sh}(\sqrt{\eps_0^2+\eps_B^2}\,\beta)\cdot\frac{\eps_B}{\sqrt{\eps_0^2+\eps_B^2}}
-{\rm sh}(\eps_B\,\beta)}
{{\rm ch}(\sqrt{\eps_0^2+\eps_B^2}\,\beta)-{\rm ch}(\eps_B\,\beta)}
\,\,.
\eeq
This formula can be evaluated in various limits, i.e.\ due to the smallness
of $g^*$ for GaAs, one can consider for example the limit
$\frac{g^*\eps_B\,\beta}{2}\ll 1$ while at the same time $\eps_B\beta\gg 1$
(i.e.\ $\frac{\hbar|e|B}{2k_BT}\gg{k_BT}$), which is somewhat strange,
although not unreasonable, if one considers fields in the Tesla range and
temperatures in the MilliKelvin region.

In the following, we also consider {\it adiabatic} demagnetization or
magnetization processes, i.e.\ where during the change of $B$ and the
ensuing measuring processes the {\it entropy} of
the dot is kept constant. This only implies  that the changes of the
$B$-field, and the measuring processes considered, must be much faster than
the thermal relaxation of the dot to the surroundings, which is not
unreasonable, since with advanced techniques magnetic fields can at present be
changed significantly  in two picoseconds, \cite{refBack,refLeineweber},
 whereas the thermal relaxation of the
electronic state of a quantum dot can be  much slower, i.e.\ by several
orders of magnitude, \cite{refAwschalom}.
 
Now an adiabatic change $\Delta B$ leads to a corresponding change $\Delta
T$, which is given by the relation

\beq\label{eqVonsovskij}
\left (\frac
{{\rm d} T}{{\rm d} B}\right )_S =
 -\frac{
\left (\frac{\partial S}{\partial B}\right )_T
}{
\left (\frac{\partial S}{\partial T}\right )_B}
=-\frac{T}{C_B}\left (
\frac{\partial M}{\partial T}
\right )_B=+\frac{\beta}{C_B}\left (
\frac{\partial M}{\partial \beta}
\right )_B\,.
\eeq
Here we have used the well-known relations $\frac{\pa S}{\pa B}=\frac{\pa
M}{\pa T}$  (which follows from ${\rm d}G=-M{\rm d}B-S{\rm d}T$) and
$C_B=T\frac{\pa S}{\pa T}$ (the 'heat capacity of the dot' at constant $B$).
So from $\frac{\pa M}{\pa\beta}$, i.e.\ from ({\ref{eqM(T,B)}), knowing
$\Delta B$ and $C_B$ (which must be $>0$ and can be calculated from the
formula $C_B(T,B)=k_BT\cdot\pa^2 [T\ln Z(T,B)]/\pa T^2$), one can directly
evaluate $\Delta T$.

So at very low temperatures, i.e.\ if $\frac{|g^*|\,\eps_B\,\beta}{2}\gg 1$ and
-- of course -- $\eps_0\beta\gg 1$, one obtains with the
relation ${\rm th}(x)\cong 1 - 2\cdot e^{-2x}+...$, valid asymptotically for
$x\gg 1$:
 \beq\label{eqdM/dT}
\frac{\pa\tilde M}{\pa\beta}:=\frac{1}{\mu_B^*}\,\frac{\pa M(T.B)}{\pa
\beta}\cong\eps_B\cdot[(g^*)^2\,e^{-|g^*|\,\eps_B\,\beta}-2\,e^{-2\eps_0\,\beta}]
\,\,.
\eeq
But the heat capacity $C_B$ should remain positive for finite $T$.
So for GaAs, in a range of sufficiently low temperatures and sufficiently
low fields, i.e.\ for temperatues $T$ {\it below} ({\it above}) a  value $T_0(B)$ given
by $\exp\left [-\frac{2\eps_0-|g^*|\eps_B}{k_B\,\cdot T_0(B)}\right ]=\frac{(g^*)^2}{2}$, adiabatic
demagnetization (${\rm d}B < 0$) leads to a {\it decrease} ({\it increase})
of $T$.
 If -- on the other hand -- we do {\it not} assume $|g^*|\eps_B\beta \gg 1$,
but the opposite limit $|g^*|\eps_B\beta\ll 1$, then we obtain
$\frac{\pa\tilde M}{\pa\beta}=\eps_B\cdot
[\frac{(g^*)^2}{4}-\frac{k_BT}{\eps_0}]$, leading to a similar conclusion,
now with $k_BT_0(B)\cong\eps_0\cdot\frac{(g^*)^2}{4}$, not depending on $B$.

In Fig.\ 1, for various values of $B$, we plot the values of $\tilde
M(T,B):=\frac{M(T,B)}{\mu_B^*}$ against the temperature $T$ -- ranging
from 0
K to $\approx 0.008$ K -- and the magnetic induction $B$ -- ranging from 0 to
$0.08$ Tesla; the
characteristic line $T_0(B)$ separating positive and negative values of $M$ is
given by the  third-lowest contour line from the right, which
 ends for $T_0\to 0$ at a value $B_k\approx 0.048$ Tesla, and for $B\to 0$ at a value
$T_k\approx 0.008$ K.

{\it These are the values for GaAs, calculated with $\eps_0 =3.37$ meV.}
({\it The
corresponding values for
$\eps_0=7.5$ {\rm meV} are: $T_k\approx 0.018$ {\rm K}; $B_k\approx 0.1$
{\rm Tesla}, i.e.\ they
scale roughly $\sim \eps_0$, as expected.})

In Fig.\ 2, the {\it adiabatic derivative} $(\frac{{\rm d}T}{{\rm
d}B})_S$ from eq.\ (\ref{eqVonsovskij}) is plotted over $B$ rsp.\ $T$ ranging from
0 to 5 Tesla rsp.\ from 0 to 6 K in a 3d-representation with contour lines.
The special contour line separating positive and negative values of $(\frac{{\rm d}T}{{\rm
d}B})_S$
 does hardly depend on $B$ over an extremely wide range of $B$-values,
and is clearly visible (it is the line vertically above the points with
$T\approx 3\,\,{\rm K}$).      In agreement with
the '3rd principal law of thermodynamics', the adiabatic derivative
$(\frac{{\rm d}T}{{\rm d}B})_S$ always vanishes for $T\to 0$, for all
values of $B$. But one should note that according to Fig.\ 2 and Fig.\ 3,
 $(\frac{{\rm d}T}{{\rm d}B})_S$
piles up to very high values in the region  $ 1 \,\,{\rm} K\,\,\lsim T\,\, \lsim
\,\,1.8
\,\,{\rm K}$, for  $B$-values $\lsim 0.016$ Tesla:  Namely, as seen in Fig.\
3, in this 'sensitive region' one can easily obtain values of the adiabatic derivative
between 100 and 500, and even larger values for temperatures around 1.5 K,
if the external magnetic field is around 0.001 Tesla. Note, however, that in our
theory we cannot consider naively the limit $B\to 0$, since the
characteristic magnetic lengh, the 'cyclotron radius'
$l_m(B):=\sqrt{\frac{\hbar}{m^*\cdot\om_c}}=\sqrt{\frac{\hbar c}{e\cdot B}}$,
should be much smaller than the distance of two dots, or much smaller than
any other geometrical extension of our 2d GaAs dot system (for $B=1$ Tesla,
$l_m$ is 25.7 nm).  Keeping this constraint in mind, concerning the
change of the temperature by an infinitesimal adiabatic demagnetization in
the above-mentioned 'sensitive region', we have the following result:

 For an isolated qantum dot in 2d-GaAs, with N=1 electrons on the dot,
starting at the point ($T\approx 1.5$ K, $B\approx 0.01$ Tesla), for $\hbar
\omega_0 \approx 3.37$ meV, we get
$ \Delta T \,\,{\rm [K]} \rsim 100\cdot \Delta B \,\,{\rm [Tesla]}
\,$. This implies that  an unusually small adiabatic change of the magnetic
field can lead to a significant change of the electron temperature in the
dot, if one roughly hits the above-mentioned region.
 
{\it Thus, on the one hand, we have the change of
sign of the adiabatic derivative from positive values for $T<3$ K to negative
values for $T>3$ K, a remarkable phenomenon in itself. On the other hand we
have the fact that  the change $\Delta T$  in the
         'sensitive region' is unusually large also in magnitude, i.e.\
there it is really important to explicitly consider also the spin, and not
only the orbital motion.}

Therefore, to diminish the dot temperature (compared with the surrounding solid)
e.g.\ by $\Delta T=-0.1$ K (at least for a transient time $\tau_T$, which is
determined by the small coupling of the dot to the degrees of freedom of the
surrounding system, and which we assume to be much larger than the time
$\tau_B$ necessary for significant changes
$\Delta B$ of the magnetic field), in the above-mentioned region it is only
 necessary to decrease the magnetic
field  by $B\lsim  10^{-3}$ Tesla.

 After having reached the thermodynamic equilibrium of the dot with its
surroundings, its temperature increases again, but that of the solid decreases,
until they equalize, i.e.\ the final temperature has been lowered in any
case. After that, the magnetic field can be turned off {\it
isothermally} and the process can be periodically iterated. In such a way
this process can be used for {\it magnetic cooling} of the dot system, which
gives a flavour of the 'new physics' involved by controlling the magnetism
of the dot.

 All this will be considerably more effective -- and also more interesting
-- for a large number of dots and for $N\gg 1$ electrons per dot~:
\section{The case $N\gg 1$ -- a simplified Hartree aproach:}
\subsection{Basic approximations}
If the dot contains $N$ electrons, the  Hamiltonian is
\beqa\label{eq11}
\hat{\cal H}&=&\sum\limits_{j=1}^N\,\left
\{-\frac{\hbar^2}{2m^*}\,\nabla^2_j+\frac{1}{2}m^*\omega^2r_j^2
+\frac{\hbar\omega_c}{2}\cdot\left (\hat l_{z,j}+g^*\,\hat S_{z,j} \right )
\right \}\cr
&+&\frac{1}{2}\sum\limits_{i,j}^{(\ne)}\,\frac{e^2}{\epsilon\,|\vec r_i-\vec
r_j|}\,.
\eeqa
Here the inclusion of the Coulomb interaction for our multi-electron planar
parabolic quantum dot leads to considerable complications in comparison to
free particles, since the Coulomb energy is of the same order of magnitude
as the kinetic energy for electrons confined in dots. This Coulomb
interaction is known to consist of the {\it direct} term (Hartree term) and
the {\it exchange} term (Fock term): The former interaction is of long-range
type, while the second one is short-ranged (cf.\
\cite{Maksym}-\cite{Maksym2}) and oscillatory in it's
position-dependence, cf.\ eq.\  (13) in
\cite{Chamon}. As a consequence, as shown in a long calculation in
\cite{refJacak97} for which we do not have a shorter argument, the ratio
of the exchange energy divided by the Hartree energy decreases in d=2
dimensions as $N^{-1/4}$.
Therefore at sufficiently high $N$  $(\rsim 10^2$--$10^3$) the
exchange  should no longer play the usual all-important central role --
considering also exact calculations for ${\cal O}(10)$ electrons,
see e.g.\ \cite{Chamon}, which show
that then the above-mentioned ratio is $\lsim 1/3$.
  So we neglect the exchange in a kind of zeroth-order approximation which still
gives interesting analytical results for the $B$- and $T$-dependence
 (see below) generalizing directly
 those of the preceding chapter 2. (Including the exchange would
 preclude this analysis.)

Thus, from (\ref{eq11}), we arrive at  Hartree equations of the form, \cite{refLandau}:
\beq\label{eq12}
\left \{
-\frac{\hbar^2}{2m^*}\nabla^2+\frac{1}{2}m^*\omega^2r^2
+\frac{\hbar\omega_c}{2}\left (\hat l_z+g^*\hat S_z\right )
+V_j(\vec r)
\right \}\psi_{p_j}(q)\,=\,\epsilon_{p_j}\psi_{p_j}(q)\,,
\eeq
where $q :=(\vec r,\sigma)$; $j=1,2,...,N$, and where the lower index $p_j$
represents a triple of the {\it three} quantum numbers $n_+$, $n_-$, and
$s_z$, and where $V_j(\vec r)$ has to be determined self-consistently~:
\beqa V_j(\vec r) &:=& \frac{e^2}{\epsilon}\,\int\,{\rm d^2}r^{\,'}\,\frac{n_j(\vec
r^{\,'})}{|\vec r -\vec r^{\,'}|}\,,\cr
n_j(\vec r) &:=& \sum\limits_{i(\ne j)=1}^N\,\sum\limits_\sigma
\,|\psi_{p_i}(\vec r,\sigma)|^2\,.\nonumber
\eeqa
$\epsilon$ is the dielectric constant of the solid, e.g.\ $\epsilon \approx
12.5$ for GaAs.

(Note that a numerical calculation supports 
 the application of the
Hartree approach, at least for a qualitative behaviour of multi-electron
dots, \cite{Wagner}.) 

For a solution of the Hartree equations we use as  zeroth approximation
the semi-classical formula for $n(r)$ given in \cite{refShikin}, i.e.:
\beq \label{eqDemel}
n_j(\vec r)\approx n(\vec
r)=\cases{\frac{3N}{2\pi R^2}\sqrt{1-\frac{r^2}{R^2}} &for $r\le R$\cr
0 &else\,.\cr} \,.
 \eeq
This approximate formula, which admittedly contradicts the boundary
conditions for the harmonic oscillator functions, describes at least
qualitatively the density of electrons inside the dot (comparison with exact
numerical results for a small number of confined electrons,
\cite{Wagner,Pfannkuche}, shows that the quantum corrections to $n(r)$ do
not modify it essentially except near the edge in cases of 'edge
reconstruction', as discussed in the already mentioned paper \cite{Chamon},
see also chapter 4.7 in
\cite{refJacak}).

As usual, for parabolic confinement we rely  on the close connection between
the 'effective dot radius $R$' used in (\ref{eqDemel}) and the
multi-electron wavefunction; therefore one can consider $R$ also as the
fitting parameter for this wavefunction.

In zeroth order of perturbation we have with (\ref{eqDemel})
for small $r/R$~:
\beq
V_j(r)\approx \frac{e^2}{\epsilon}\int\,{\rm d}^2r'\,\frac{n(\vec
r^{\,'})}{|\vec r -\vec r^{\,'}|}\approx\frac{3\pi Ne^2}{4\epsilon R}\cdot
\left (1-\frac{r^2}{2R^2}\right )\,,
\eeq
  As a consequence, in the interior of the dot (and not only there, see the
 remark below) we now get for the Hartree differential equation the
 simple 'renormalized form'~:
\beq\label{eq13}
\left \{
-\frac{\hbar^2}{2m^*}\nabla^2+\frac{1}{2}m^*\Omega^2r^2
+\frac{\hbar\omega_c}{2}\left (\hat l_z+g^*\hat S_z\right )
\right \}\psi_{p}(q)\,=\,E_{p}\psi_{p}(q)\,,
\eeq
i.e.\ through this equation we now have an {\it effective single-particle
equation}, 
where the 'renormalized confining frequency' $\Omega$, the 'renormalized
single-particle energy' $E_p$, and the 'renormalized cyclotron length' $l$ are
defined as
\beq\label{eq14}
\Omega^2 := \omega^2-\frac{3\pi N e^2}{4\epsilon m^* R^3};\quad
E_p := \epsilon_p -\frac{3\pi Ne^2}{4\epsilon R};\quad
l^2 := \frac{\hbar}{m^*\Omega}\quad .
\eeq
So all three quantities are now $R$-dependent, which should be kept in mind.

Note that for our case, i.e.\ for $N\gg 1$, one has $R\gg l$; so almost all
single-particle wave functions should  be exponentially small for
$r\approx R$;  therefore only a neglegible number of electrons is located
near the edge of the dot, and the solutions of eq.\ (\ref{eq13}) given by Eqs.\
(\ref{eq3}) and (\ref{eq4}) can also be applied for $r>R$.
\subsection{Thermodynamics}
At $T> 0$ the probability to find an electron of the dot in a state with
an energy $\epsilon_p=\epsilon_{n_+,n_-,s_z}+\frac{3\pi Ne^2}{4\epsilon R}$,
i.e.\ $E_p=\epsilon_{n_+,n_-,s_z}$,
is defined by the Fermi distribution $n_s(\eps)$, i.e.
\beq\label{eq16}
n_s(\epsilon_+ n_++\epsilon_-n_-) :=\left [1+\exp\left
(\frac{\eps_+n_++\eps_-n_--\mu_s}{k_BT}\right )\right ]^{-1}\,,
\eeq
where $\mu =\mu(T,N,B)$ is the chemical potential of the electrons in the
dot, and with $s=s_z=\pm\frac{1}{2}$
\beq\label{eq17}
 \mu_{s_z} := \mu -\frac{\eps_++\eps_-}{2}
-g^*\frac{\om_c}{2}s_z-\frac{3\pi Ne^2}{4\eps R}\,;
\quad
\eps_\pm :=\hbar\Omega\pm \frac{1}{2}\hbar\om_c\,\,.
\eeq
The chemical potential $\mu$ can be determined  as usual from the  condition that
$N =\sum\limits_{s=-\frac{1}{2}}^{+\frac{1}{2}}\,N_s$, with
\beq\label{eq18}
N_s(T,B,\mu ) = \sum\limits_{n_+=0}^\infty\sum\limits_{n_-=0}^\infty \,n_s(\eps_+n_+
+\eps_-n_-)\,.
\eeq
Let us now introduce the {\it Grand Thermodynamic Potential} ${\cal
Y}(T,B,\mu)=\sum\limits_{s=-\frac{1}{2}}^{+\frac{1}{2}}\,{\cal Y}_s(T,B,\mu)$,
  through
\beqa\label{eq19}
{\cal Y}_s
&:=&\sum\limits_{n_+,n_-=0}^\infty\,\phi_s(\eps_+n_++\eps_-n_-)\,,\quad{{\rm
with}}\cr
\phi_s(\eps ) &:=&\,-k_BT\cdot\ln 
\left [
1+\exp \left (\frac{\mu_s-\eps
}{k_BT}\right )_{N,B}  
\right ]
\,\,.
\eeqa
Then the 'mean   energy' $E(T,B,N)$ (i.e.\ the {\it internal enthalpy},
 \cite{refREM1}), of the dot), and its {\it Free Enthalpy} $G(T,B,N)=E-T\cdot
 S$, where $S$ is the entropy, are determined by the equations

\beq\label{eq20}
E={\cal Y}+\mu N -T\left ( \frac{\pa{\cal Y}}{\pa T}\right )
-\frac{9\pi}{20}\cdot\frac{N^2e^2}{\eps R}\,,
\eeq
\beq\label{eq21}
G={\cal Y}+\mu N-\frac{9\pi}{20}\cdot\frac{N^2e^2}{\eps R}\,.
\eeq
Here the final term in eqs.\ (\ref{eq20}) and (\ref{eq21}) represents the
'double counting correction' of the Coulomb energy, where we have used that
$$
\frac{e^2}{2\eps }\int\,{\rm d}^2r\,\int{\rm d}^2
r'\,\frac{n(r)\,n(r')}{|\vr-\vr^{\,'}|}=
\frac{3\pi}{10}\cdot\frac{N^2e^2}{\eps R}\,.
$$
At $T\to 0$,  $E$ and $G$ transform into the energy of the ground state of
the dot: $E_0 =\lim\limits_{T\to 0}E=\lim\limits_{T\to 0}G={\cal
Y}_0+\eps_FN-\frac{9\pi}{20}\cdot\frac{N^2e^2}{\eps R_0}$, where ${\cal
Y}_0 :=\lim\limits_{T\to 0}\,{\cal Y}$; $\eps_F :=\lim\limits_{T\to
0}\,\mu$; $R_0 :=\lim\limits_{T\to 0}\,R$. 

Now for finite temperatures we define our 'effective dot
radius' $R=R(N,B,T)$  (or more precise: the 'effective radius of the electron
liquid on the dot')
from the condition that the Free Enthalpy should fulfill
\beq\label{eq22}
\left (\frac{\pa G}{\pa R}\right )_{|N,T,B} \,=\,0\,\,.
\eeq
Again, this condition couples spin and orbital degrees of freedom.

As in \cite{refJacak97}, we now use  {\it Laplace transforms} of the
quantities appearing in (\ref{eq18}) and (\ref{eq19}), marked by a 'tilde';
e.g.\ we write
$$
n_s(\eps )=\frac{1}{2\pi{\rm i}}\int\limits_{c-{\rm i}\infty}^{c+{\rm
i}\infty}\,{\rm d}p\,\,\tilde n_s(p)\cdot e^{+p\cdot\eps }\quad ,{\rm\,\, with}\quad
\tilde n_s(p)=
\int\limits_0^\infty\,{\rm d}\eps\,n_s(\eps )\cdot e^{-p\cdot\eps }\,,$$
where $c$ is an arbitrary real number, which must only be 'positive enough'
to ensure existence of the transformation  (see below).
Then $N_s$ and ${\cal Y}_s$ can be represented in the following form:
\beq\label{eq23}
N_s=\int\limits_0^\infty \,\left ( -\frac{\pa n_s(\eps )}{\pa \eps} \right )
\cdot Z(\eps )\,{\rm d}\eps \,\,,
\eeq 
\beq\label{eq24}
{\cal Y}_s = -\int\limits_0^\infty\,n_s(\eps )\,Z(\eps)\,{\rm d}\eps\,,
\eeq
where the Laplace transform of $Z(\eps )$ is given by the simple expression
\beq\label{eq25}
\tilde Z(p) =\frac{1}{p\cdot (1-e^{-\eps_+\,p})\cdot(1-e^{-\eps_-\,p})}\,.
\eeq
Here the constant $c$ in the Laplace transform (see above) has to be chosen
in such a way that all singularities of $\tilde Z(p)$ are situated to the
left of the straight line $(c-{\rm i}\infty ,c+{\rm i}\infty )$ -- this gives
a precise meaning to the above-mentioned formulation 'positive enough'-- : Then the
contour of integration in (\ref{eq25}) can be closed at infinity, and we can
use the residuum calculus to evaluate the integral. It is easy to show
that $\tilde Z(p)$ has at the same time poles of first order at the points
$p=p_n^{(\pm
)}:=2\pi {\rm i}\,n\,\frac{\hbar\Om}{\eps_\pm}$, with $n=\pm 1,\pm 2,...$, and a
pole of third order at $p=0$, if the quantity $X:=\frac{\om_c}{2\Om}$ is
$\ne X_0$, where
\beq\label{eq26}
X_0\,=\,0\,,\,\frac{1}{3}\,,\,\frac{1}{5}\,,\,\frac{3}{5}\,,\,...\quad .
\eeq
If, on the other hand, the non-generic condition $X=X_0$ is fulfilled, then $\tilde
Z(p)$ has poles of first, second and third order.

From now on we will be interested only in the 'generic situation', i.e.\ in
 such fields, for which $X\ne
X_0$. Having found all the residues of $\tilde Z(p)$ and performing the
summation with respect to all poles, we find an expression for  $Z(\eps )$.
Then, substituting $Z(\eps )$ into the integrals (\ref{eq23}) and
(\ref{eq24}) and using the properties of the Fermi functions, we find 
expressions for $N$ and ${\cal Y}$. In the low-temperature limit $k_BT\ll
\mu_s$, these quantities take the form
\beqa\label{eq27}
N&=&\frac{\mu^2}{\eps_0^2}-\frac{1}{2}+((g^*)^2-1)\cdot\left
(\frac{\hbar\om_c}{2\eps_0}\right )^2 +\frac{\pi^2}{3}\cdot\left
(\frac{k_BT}{\eps_0}\right )^2\cr
&+&\sum\limits_{s=-\frac{1}{2}}^{+\frac{1}{2}}\left \{ P_1^{(+)}\left
(\frac{\mu_s\eps_+}{\eps_0^2}\right )-\frac{\eps_-}{\eps_+}\cdot P_2^{(+)}
\left (\frac{\mu_s\eps_+}{\eps_0^2}\right )\right \}\nonumber\\
&+&\sum\limits_{s=-\frac{1}{2}}^{+\frac{1}{2}}\left \{
P_1^{(-)}\left(\frac{\mu_s\eps_+}{\eps_0^2}\right )
-\frac{\eps_+}{\eps_-}\cdot P_2^{(-)}
\left (\frac{\mu_s\eps_-}{\eps_0^2}\right )
 \right \}\,,
\eeqa
\beqa\label{eq28}
{\cal Y}&=&-\frac{1}{3}\frac{\mu^3}{\eps_0^2}+\frac{\mu}{2}-\mu
\cdot((g^*)^2-1)\cdot\left
(\frac{\hbar\om_c}{2\eps_0}\right )^2 -\frac{\hbar\Om}{2}\cdot
\left [1+\frac{4}{3}\left ( \frac{\hbar\om_c}{\eps_0}\right )^2\right ]
\cr
&-&\mu\cdot\frac{\pi^2}{3}\cdot\left (\frac{k_BT}{\eps_0}\right )^2
+\eps_-\sum\limits_{s=-\frac{1}{2}}^{+\frac{1}{2}}\left \{ P_1^{(+)}\left
(\frac{\mu_s\eps_+}{\eps_0^2}\right )+\frac{\eps_-}{\eps_+}\cdot P_2^{(+)}
\left (\frac{\mu_s\eps_+}{\eps_0^2}\right )\right \}\nonumber\\
&+&\eps_+\sum\limits_{s=-\frac{1}{2}}^{+\frac{1}{2}}\left \{
P_1^{(-)}\left(\frac{\mu_s\eps_+}{\eps_0^2}\right )
+\frac{\eps_+}{\eps_-}\cdot P_2^{(-)}
\left (\frac{\mu_s\eps_-}{\eps_0^2}\right )
 \right \}\,,
\eeqa
where $\eps_0^2:=\hbar^2\cdot \left (\om_0^2-\frac{3\pi}{4}\frac{Ne^2}{\eps
\,m^*R^3}\right )$, which corresponds to the first equation in (\ref{eq14}).
Here the periodic functions $P_m^{(\pm )}(z)$ have for {\it even} $m$ the form
\beq\label{eq29}
P_{m}^{(\pm)}(z) =\sum\limits_{n=1}^\infty\frac{2\pi^2nkT\eps_\pm}
{\eps_0^2}\cdot\left [{\rm sh}\left ( \frac{2\pi^2 nkT\eps_\pm}
{\eps_0^2}\right )\right ]^{-1}\cdot\frac{\cos(2\pi nz)}{2^{m-1}\pi^{m}n^{m}}\,,
\eeq
whereas for {\it odd} $m$ the same result applies, if the functions
$\cos(2\pi nz)$ in
(\ref{eq29}) are replaced by $\sin(2\pi n z)$.

Finally, we obtain the following expression for the Free Enthalpy~:
\beqa\label{eq30}
G=\frac{2\mu^3}{3\eps_0^2}+\frac{9\pi N^2e^2}{20\,\eps\,R}
+\mu\sum\limits_{s=-\frac{1}{2}}\left \{
P_1^{(+)}\left ( \frac{\mu_s\eps_+}{\eps_0^2}\right )
+P_1^{(-)}\left ( \frac{\mu_s\eps_-}{\eps_0^2}\right )\right \}\cr
-\mu\sum\limits_{s=-\frac{1}{2}}\left \{
\frac{\eps_-}{\eps_+}P_2^{(+)}\left ( \frac{\mu_s\eps_+}{\eps_0^2}\right )
+\frac{\eps_-}{\eps_+}P_2^{(+)}\left ( \frac{\mu_s\eps_+}{\eps_0^2}\right )
\right \}\cr
+\sum\limits_{s=-\frac{1}{2}}^{+\frac{1}{2}}\left \{
\eps_-P_2^{(+)}\left (\frac{\mu_s\eps_+}{\eps_0^2}\right )
+\eps_+P_2^{(-)}\left (\frac{\mu_s\eps_-}{\eps_0^2}\right )
\right \}\cr
+\sum\limits_{s=-\frac{1}{2}}^{+\frac{1}{2}}\left \{
\frac{\eps_-^2}{\eps_+}P_3^{(+)}\left (\frac{\mu_s\eps_+}{\eps_0^2}\right )
+\frac{\eps_+^2}{\eps_-}P_3^{(-)}\left (\frac{\mu_s\eps_-}{\eps_0^2}\right )
\right \}\cr
-\frac{\hbar\Om}{2}\cdot \left [ 1+ \frac{4}{3}\cdot \left
(\frac{\hbar\om_c}{2\eps_0}\right )^2\right ]\,.
\eeqa
The chemical potential $\mu$ is found from (\ref{eq27}), which can be rewritten as
\beqa
\frac{\mu^2}{N\eps_0^2} &=&1 +\frac{1}{N}\left \{
\frac{1}{2} -\left ((g^*)^2-1 \right )\cdot \left (\frac{\hbar\om_c}{2\eps_0} \right )^2
-\frac{\pi^2}{3}\cdot \left (\frac{k_BT}{\eps_0}\right )^2
 \right \}\cr
&-&\frac{1}{N} \left \{
\sum\limits_{s=-\frac{1}{2}}^{\frac{1}{2}}\left [
P_1^{(+)}\left ( \frac{\mu_s\eps_+}{\eps_0^2}\right )
+P_1^{(-)}\left ( \frac{\mu_s\eps_-}{\eps_0^2}\right )\right ]\right \}\cr
&-&\frac{1}{N} \left \{
\sum\limits_{s=-\frac{1}{2}}^{\frac{1}{2}}\left [
\frac{\eps_-}{\eps_+}P_2^{(+)}\left ( \frac{\mu_s\eps_+}{\eps_0^2}\right )
+\frac{\eps_-}{\eps_+}P_2^{(+)}\left ( \frac{\mu_s\eps_+}{\eps_0^2}\right )
\right ]
\right \}\nonumber
\eeqa
In the lowest approximation, if the number of electrons in the dot is large
($N\gg 1$), then at moderately low magnetic fields ($\hbar\om_c \ll
2\eps_0$) and moderately low temperatures this expression takes the form (in
zeroth approximation,
$\mu\approx \mu_0$):
$\frac{\mu_0^2}{N\eps_0^2}=1$, or $\mu\approx\mu_0=\eps_0\sqrt{N}$.

In the next approximation, within an accuracy of order $1/\sqrt{N}$, the
chemical potential is equal to
\beq\label{eq31}
\mu \cong\eps_0\sqrt{N}\left \{ 1 +\frac{1}{2N}f(N,B,T)\right \}\,,
\eeq
where the quantity
\beqa
f(N,B,T) &=&
\frac{1}{2} -\left ((g^*)^2-1 \right )\cdot \left (\frac{\hbar\om_c}{2\eps_0} \right )^2
-\frac{\pi^2}{3}\cdot \left (\frac{k_BT}{\eps_0}\right )^2
 \cr
&-&
\sum\limits_{s=-\frac{1}{2}}^{\frac{1}{2}}\left [
P_1^{(+)}\left ( \frac{\mu_s\eps_+}{\eps_0^2}\right )
+P_1^{(-)}\left ( \frac{\mu_s\eps_-}{\eps_0^2}\right )\right ]\cr
&-&
\sum\limits_{s=-\frac{1}{2}}^{\frac{1}{2}}\left [
\frac{\eps_-}{\eps_+}P_2^{(+)}\left ( \frac{\mu_s\eps_+}{\eps_0^2}\right )
+\frac{\eps_-}{\eps_+}P_2^{(+)}\left ( \frac{\mu_s\eps_+}{\eps_0^2}\right )
\right ]
\nonumber
\eeqa
is of order-of-magnitude ${\cal O}(1)$, and from (\ref{eq17}) one gets
$$\mu_s\approx (\mu_0)_{s_z}=\eps_0\sqrt{N}+\sqrt{\eps_0^2+\left
(\frac{\hbar\om_c}{2}\right )^2}
-g^*\frac{\hbar\om_c}{2}s_z-\frac{3\pi Ne^2}{4\eps R}\,.$$
If we substitute this  expression for $\mu(N,B,T)$ in (\ref{eq30}), we
obtain the Free Enthalpy as a function of $N$, $B$ and $T$. As follows from 
(\ref{eq31}), the first summand in the r.h.s.\ of (\ref{eq30}) is of order
$N^{\frac{3}{2}-2\sigma}$, if $\eps_0\sim N^\sigma$ (it can be shown 
that $\sigma =1/6$, see eq.\ (\ref{eq33}) below). The second summand  on the r.h.s.\ of (\ref{eq30}) is
of order $N^{2-\gamma}$ (with $\gamma=\frac{1}{3}$, see (\ref{eq33})). All other
summands have still less
order of magnitude. If we retain only the first two terms in the r.h.s.\ of
(\ref{eq30}) and use (\ref{eq22}), then we come to the following equation
for $R$, \cite{refJacak97}:
\beq\label{eq32}
\om_0^2 \cong \left (\frac{3\pi}{4}\cdot \frac{Ne^2}{\eps\,R^3m^*}\right )\cdot 
\left \{1+\frac{100\, a_B^*}{27\pi R}\right \}\,,
\eeq
where $a_B^*:=\frac{\hbar^2\eps}{m^*e^2}$ is the {\it effective Bohr
radius}. 

This means that $\om_0 \,\,(=\frac{\eps_0}{\hbar})$ is essentially
identified with the plasma frequency calculated from the electron density in
the dot calculated with $T=0$ and $B=0$;
 in this approximation the radius $R$ of the dot does not depend on
$B$ and $T$ at all and is defined only by the number $N$ of electrons on the
dot. To obtain a dependence of $R$ on $B$ and $T$ while solving equ.\
(\ref{eq32}), it is in principle necessary to take into account corrections
of higher order; yet at $N\gg 1$ the corrections are very small, namely of
relative order ${\cal O}(N^{-\frac{1}{2}})$, and we neglect them.

 For $a_B^*\ll R$ the solution of $(\ref{eq32})$ in the first
approximation is
\beq\label{eq33}
R\cong R_0\cdot \left (1+\frac{100\,a_B^*}{51\pi R_0}\right )\,,
\eeq where $R_0 :=(\frac{3\pi Ne^2}{4\eps\,m^*\om_0^2} )^{\frac{1}{3}}$.
\newline Hence
$\eps_0 = \hbar\cdot\left [ \om_0^2-\frac{3\pi N e^2}{4\eps m^*R^3}\right
]^{\frac{1}{2}}=\hbar\om_0\cdot\left [{(\frac{100 a_B^*}{27\pi
R}})/{(1+(\frac{100 a_B^*}{27\pi R}) }) \right ]^{\frac{1}{2}}\propto
R^{-\frac{1}{2}}\propto N^{-\frac{1}{6}}$, and $\frac{9\pi^2N^2e^2}{20\eps
R}\propto \frac {N^2}{R}\propto N^{2-\frac{1}{3}}$, as already stated above.

From Eq.\ (\ref{eq32}) it follows that $R$ transforms into $R_0$ in the
classical limit ($\hbar\to 0$), cf.\ Eq.\ (\ref{eqDemel}). Moreover, the
dependence of $R$ with respect to $N$ and $\hbar\omega_0$ as above,
corresponds well to numerical results from \cite{Maksym2}.

 Now the magnetic moment of a dot is defined by the derivative
$$ M=-\left (\frac{\pa{\cal Y}}{\pa B}\right )_{\mu ,T}\,\,.$$
Using expression (\ref{eq29}) for ${\cal Y}$ we find $M$, taking into
account
terms of order $1/\sqrt{N}$ :
\beq\label{eq34}
\frac{M}{\mu_B^*} = \sqrt{N}\cdot
\left [ 
\left ((g^*)^2-1\right
)\cdot\frac{\mu_B^*\cdot B}{\eps_0}
+2\frac{P_1^{(+)}\left (\frac{\mu_0\eps_+}{\eps_0^2}\right )
-P_1^{(-)}\left (\frac{\mu_0\eps_-}{\eps_0^2}\right
)}{\sqrt{1+\frac{(\mu_B^*)^2\cdot B^2}{\eps_0^2}}}
\right ]
\,,
\eeq
where $\mu_0=\eps_0\sqrt{N}$ has been defined above.

We see from this expression that $M\to 0$ for $B\to 0$, since then
$\eps_+=\eps_-$. At the same time the
first summand on the r.h.s.\ of (\ref{eq34}) is {\it negative} ($(g^*)^2 <
1$) and {\it monotoneously decreasing} with increasing $B$. However the
second summand {\it oscillates} around zero with increasing $B$, and so the
possibility
 exists that
$\frac{1}{B}\frac{M}{\mu_B^*}$ may be {\it positive} at $B\to 0$. Let us find out conditions
when this case can happen:

 For this we expand the r.h.s.\ of (\ref{eq34}) into a series in terms of
$B$, and in a linear approximation in $B$ we get for
$N\gg 1$ :
\beqa\label{eq35}
\frac{M}{\mu_B^*} &=&\sqrt{N}\cdot \left ((g^*)^2-1\right )\cdot
\frac{\mu_B^*\cdot B}{\eps_0}
+4N\cdot P_0(\sqrt{N})\cdot\frac{\mu_B^*\cdot B}{\eps_0}\cr
&\approx& 4N\cdot P_0(\sqrt{N})\cdot\frac{\mu_B^*\cdot B}{\eps_0}\,,
\eeqa

The function $P_0(x)$ is periodic, with period $\Delta x=1$, namely
\beq\label{eq36}
P_0(x) = 2\sum\limits_{n=1}^\infty\,\frac{2\pi^2 nk_BT}{\eps_0}\cdot\left
[{\rm
sh}\left (\frac{2\pi^2 nk_BT}{\eps_0}\right )\right ]^{-1}\cdot \cos (2\pi
nx)\,,    
\eeq
which can also be written as
$P_0(x)
=\left \{\sum\limits_{n=-\infty}^{+\infty}\, A(x-n)\right \} -1 \,\,, 
$ where the functions $A(x-n)$ are calculated in the Appendix.
As a consequence  of the large-$n$-behaviour of the Fourier coefficients in
front of $\cos (2\pi n x)$ in (\ref{eq36}),
 the function $A(x)$ reaches a sharp maximum at $x=0$, 
namely~:
$$ A(x)\equiv\frac{\eps_0}{4k_BT}\cdot\left [{\rm ch}\left
(\frac{\eps_0x}{2k_BT}\right )\right ]^{-2} \to A(0)=\frac{\eps_0}{4k_BT}\,.
$$
{Thus the function $P_0(\sqrt{N})$ in eq.\ $($\ref{eq35}$)$ takes positive values
at $N=m^2$ $($$ m=0, 1, 2,...)$, if at the same time $\frac{\eps_0}{4k_BT}
>1$ $($see also eq.\ $($\ref{eq37}$)$ below$)$.}

 {\it Therefore at low fields the quantity $\frac{M}{B\mu_B^*}$ is
positive, if the following two conditions are simultaneously fulfilled, namely 
(i) the temperature has to be low enough: $T < T_0:=\frac{\eps_0}{4k_B}$,
and (ii) the number $N$ of electrons in the dot is equal to $N=m^2$, with
$m=0,1,2,...$  ${\rm (}$which corresponds for $B=0$ to a 'half filled outer shell'
condition, as we shall see${\rm )}$}.

In fact, for a planar parabolic dot in the absence of magnetic field $B$ and
in a one-electron approximation without interaction, the energy levels
of the electron are defined by quantum numbers $n=0,1,2,...$, 
($=n_++n_-$ in Eq. (\ref{eq4})) and are
degenerate with respect to the numbers 
$$ n_+-n_-= \cases{0; \pm 2; \pm 4;...,\pm n &for even $n$\cr
             \pm 1;\pm 3;\pm 5;...;\pm n &for odd $n$}\,,
$$
and, of course, also degenerate with respect to the spin.

Electronic states of a planar dot with the same quantum number $n$ form a
'shell': Then at $N=m^2$ the last electron shell (i.e.\ with the highest
possible
$n$) of the parabolic dot turns out to be just half-filled.

(Here it should be noted that in eq.\ 
(\ref{eq35}) the explicit spin dependence -- i.e.\ the term involving
 $(g^*)^2$ --
is neglegible for $N\gg 1$, and the phenomenon considered is actually
primarily a shell effect.)

As a consequence, a planar quantum dot with $N=m^2$ ($\gg 1$), at low temperatures
($T<T_0=\frac{\eps_0}{4k_B}$) and at 'sufficiently low fields' (what this means
quantitatively, is determined soon), i.e.\ for $B \to 0$,
 turns out to be a 'paramagnetic two-dimensional artificial atom'
with magnetic moment
\beq\label{eq37}
M(T,B,N)=\mu_B^*\cdot 4N\cdot \left (\frac{\eps_0}{4k_BT}-1\right )\cdot \frac{\mu_B^*\cdot
B}{\eps_0}\,\,.
\eeq
{\it Thus, when the magnetic field is increased, the magnetic moment of a
planar dot in a two-dimensional electron gas  at first increases
$\propto B $ according to (\ref{eq37}); then, according to (\ref{eq34}) it
reaches a maximum, then diminishes again, vanishes, and becomes negative.}

 Let
us first find the value $B_0$, where $M$ vanishes.
At $N=m^2 $, i.e.\ at integer $\sqrt{N}$, we have
$$ P_1^{(\pm )}\left (\frac{\mu_0\eps_\pm}{\eps_0^2}\right )
\approx P_1\left \{\sqrt{N}\cdot \left (1 \pm
\frac{\mu_B^*B}{\eps_0}\right )\right \}\equiv\pm P_1\left (\sqrt{N}\frac{\mu_B^*\cdot
B}{\eps_0}\right )\,,
$$
where $P_1(x)$ has been defined above. Then
\beq\label{eq38}
\frac{M}{\mu_B^*} = \sqrt{N}\cdot\left \{ ((g^*)^2-1)\frac{\mu_B^*\cdot
B}{\eps_0} + 4 P_1\left (\sqrt{N}\frac{\mu_B^*\cdot B}{\eps_0}\right )
 \right \}\,.
\eeq
If $T\to 0$, then at $0<x<1$ : $P_1(x) \cong \frac{1}{2} -x$. That is why at
$T=0$ for sufficiently small magnetic fields
$$
\frac {M}{\mu_B^*}=\sqrt{N}\cdot \left\{
\left ((g^*)^2-1\right )\frac{\mu_B^*\cdot B}{\eps_0}+2-4\sqrt{N}\cdot\frac{
\mu_B^*\cdot B}{\eps_0}
\right\}\,.
$$
The r.h.s.\ of this expression is positive when
\beq\label{eq39}
B < B_0 :=\frac{2\eps_0}{(4\sqrt{N}+1-(g^*)^2)\cdot\mu_B^*}\approx
\frac{\eps_0}{2\sqrt{N}\cdot\mu_B^*}\,.
\eeq
So 'sufficiently small fields' means $B\ll B_0$; i.e.\
 in  the region ($B< B_0$, $T< T_0$)
 the magnetic moment of a planar
dot is positive (i.e.\ the dot is {\it paramagnetic}). Outside this region
 $M$ is $\le 0$, and the dot is {\it diamagnetic}.

 For GaAs at
$\hbar\om_0 =3.37 $ meV we have the following typical values for $T_0$ and
$B_0$, which should
be compared with the results of Figs.\ 2,3~:
$$N=100, \quad T_0=3.26 \,\,{\rm K;}\quad B_0=0.065\,\, {\rm T}\,,$$
$$N=25, \quad T_0=4.10 \,\,{\rm K;}\quad B_0=0.164\,\, {\rm T}\,.$$

As in the case of a dot with $N=1$ electron, finally the {\it adiabatic temperature
derivative} 
 $\left ( \frac{{\rm
d}T}{{\rm d}B}\right )_S$  w.r.\ to changes of
the magnetic field is calculated through the expression

$$ 
\left ( \frac{{\rm
d}T}{{\rm d}B}\right )_S=-T\cdot \frac{\left (\frac{\pa M} 
{\pa T}\right
)_{B,N}}{C_{B,N}}\,,
$$
where $C_{B,N}$ is the heat capacity of the dot. From (\ref{eq27}) and
(\ref{eq28}) it is easy to show that
$$C_{B,N}
=\sqrt{N}\frac{2\pi^2}{3}\frac{k_B^2}{\eps_0}+{\cal{O}}(\frac{1}{\sqrt{N}})\,.$$
If we consider dots with half-filled last electron shell (i.e.\ $N=m^2$),
then according to (\ref{eq38})

$$ 
\left (\frac{\pa M}{\pa T}\right )_{B,N} = 4\mu_B^* \frac{\pa}{\pa T}
P_1\left ( \sqrt{N}\,\frac{\mu_B^*\cdot B}{\eps_0}\right )\,.
$$
(Here one should remember that $P_1(x)$ depends on $T$, see eq.\ (\ref{eq29}).)

Hence
 it follows that
\beq\label{eq40}
\left ( \frac{{\rm
d}T}{{\rm d}B}\right )_S =-4\mu_B^*\cdot\left (
\sqrt{N}\,\frac{2\pi^2}{3}\frac{k_B^2}{\eps_0}
\right )^{-1}\cdot
\frac{\pa}{\pa T}
P_1\left ( \sqrt{N}\,\frac{\mu_B^*\cdot B}{\eps_0}\right )\,.
\eeq
If we use the relationship $\frac{\pa P_1(x)}{\pa x}=P_0(x)$ and the
expression (\ref{eq36}) for $P_0(x)$, then it is possible to show that
at $0<x<1$
$$
P_1(x)=\frac{1}{2}\sum\limits_{n=-\infty}^{+\infty} \,{\rm th}\left [
\frac{\eps_0\cdot (x-n)}{2k_BT}\right ] -x+\frac{1}{2}\,.
$$
Hence it follows that the derivative $\pa P(x)/\pa T$ vanishes at x=0,
 $\frac{1}{2}$, 1. At the same time
$$
\frac{\pa ^2P_1(x)}{\pa x\pa T}_{|x=0}=\frac{\pa ^2P_1(x)}{\pa x\pa
T}_{|x=1}= -\frac{\eps_0}{4k_BT^2} < 0,\quad{\rm whereas}\quad
\frac{\pa ^2P_1(x)}{\pa x\pa T}_{|x= \frac{1}{2}} > 0\,.
$$

\noindent Thus, at low fields $(\sqrt{N} \mu_B^*\cdot B/\eps_0 \ll 1)$, similar
 as for $N=1$ (see Fig.\ 2), we get {\it positive} values of the adiabatic
temperature devative,
\beq\label{eq41}
\left ( \frac{{\rm d}T}{{\rm d}B}\right )_S = \frac{3}{2\pi^2}
\frac{\eps_0 (\mu_B^* )^2}{k_B\cdot  (k_BT )^2 }\cdot B > 0\,.
\eeq

Altogether this means that the  temperature  obtained by adiabatic
demagnetization of the dots, ${\rm d}B<0$, is strongly $B$-dependent: It
 {\it first
 diminishes with decreasing $B$}, then reaches a minimum value at
$B=B_k:=\eps_0/(2\sqrt{N}\cdot \mu_B^*)$  (since the derivative $\left ( \frac{{\rm
d}T}{{\rm d}B}\right )_S$ vanishes at
$\frac{\sqrt{N}\cdot\mu_B^*B}{\eps_0}=\frac{1}{2}$, i.e.\ $B=B_k$),
and then begins to {\it raise}, due to the inequality $\left ( \frac{{\rm
d}T}{{\rm d}B}\right )_S<0$ at $B>B_k$.

Concerning the $B$-dependence of  $\left ( \frac{{\rm d}T}{{\rm d}B}\right
)_S$, we mention -- however --  the following point:
 For $N=1$, the contour line $(\frac{{\rm
d}T}{{\rm d}B})_S=0$ in Fig.\ 2 depends on $T$ only, but not on $B$, for
 a large region of $B$ values, where it is simply given by $T\approx 3$ K.
According to Fig.\ 2, this seems only to be different for very small
$B$-values at  $T> 3$ K. In the first-mentioned respect there seems to be a
qualitative distinction between the cases $N=1$ and $N\gg
 1$, which we do not yet understand at present.

At last, using expression (\ref{eq41}), let us find out, {\it how fast} the
temperature of a dot diminishes or increases by the adiabatic change of  the magnetic
field\ : If we let in (\ref{eq41}) $T_S(B) =T +\Delta T(B)$, where $T$ is the
initial temperature, $T_S(B)$ the final temperature, and $\Delta T(B) \ll T$,
 then it is easy to show that
\beq\label{eq42}
\Delta T(B)\approx \frac{3}{4\pi^2}\cdot\left (\frac{\eps_0}{k_BT}
\right )^3\cdot
\left (\frac{\mu_B^*\cdot B}{\eps_0}
\right )^2\cdot T\,\,.
\eeq

Thus for quantum dots in a two-dimensional electron gas with  GaAs parameters
 at  a start temperature of e.g.\ $T=2$ K, and with
 a 'confining energy'  $\eps_0=\hbar \om_0
 =3.37$ meV,  if one wants to change the  electron temperature on the dot
 by $\Delta T
 =\pm 0.1$ K for $N=100$, the $B$ field has
only to change by $\pm 0.063$ T (tesla) -- for $N=25$ by $\pm 0.057$ T.\ If
$\hbar\om_0$ = 7.5 meV, then at $N=100$ the field has to change by $\pm 0.04$
 T -- at
$N=25$ by $\pm 0.035$ T.

Finally the following points should be mentioned:
as a consequence
(i) of a pronounced $B$-dependence
 of the total angular-momentum quantum numbers $L(B)$
and $S(B)$ of the ground state of the electronic ensemble and (ii)
of the Coulomb interaction of the electrons, oscillations of
the physical properties  of quantum dots with $B$ have already been
predicted and observed in a number of papers; e.g.\ even in an early
paper of Dingle, \cite{Dingle}, before the invention of quantum dots, and
later-on in papers of Maksym and Chakraborty, 
\cite{Maksym}, and in \cite{Wagner}. Furthermore,
 numerical results for {\it small} numbers of
electrons, $N={\cal O}(10)$, show that the maximum
electron density may not be at  the origin
for all electron numbers and magnetic fields (see e.g.\ 
\cite{Chamon}, \cite{Pfannkuche}, \cite{Maksym2}, \cite{Reimann}),
 and that the 'electronic edge' of the quantum dot may get
 a non-trivial structure, i.e.\ the 'edge
reconstruction', \cite{Chamon,refJacak}. However, here we
stress  for $N\gg 1$ that
\newline (i)
in principle the orbital and spin degrees of freedom are
intrinsically coupled for the individual electrons,
 although of course the total momentum quantum numbers 
$S$ and $L$ remain well-defined for circular dots, 
 and  that\newline (ii) with the quasi-classical electron density
(\ref{eqDemel}) it follows from eq.\ (\ref{eq39}) that there
 are not only periodic oscillations of the magnetization of the
dots with a period $\propto\frac{ B\cdot \sqrt{N}}{\eps_0}$, but that
 throughout these oscillations, $M$ does not remain negative
 but alternates periodically in sign.
 This is essentially a 'Hartree shell-effect', and it
seems from our analytical results that this should be
seen parallel to the exchange mechanisms, 
i.e.\ one is dealing -- so to say
-- with two different sides of one coin. This is analogous to
the situation in the conventional 3d magnetism, where both the 'Hubbard
mechanism' (in mean-field approximation essentially a Hartree effect)
 and the 'Hund's rule exchange'
are important for the magnetic properties, 
although of different relative
importance for different systems.
\section{Conclusions}
It follows from the above-stated that
\bi\item the magnetization of a planar
one-electron dot, $N=1$, see (\ref{eq6}), can be separated into two parts:
 (i) the {\it
paramagnetic} magnetization caused by the own  magnetic {\it spin moment}
 of the electron, and (ii) the orbital magnetization which is due to the
quantized {\it orbital} motion of the electron of the dot in a magnetic
field. At low temperatures ($T<T_0$) and fields ($B<B_0$) the paramagnetic
part of the susceptibility exceeds the diamagnetic one, and as a whole the
dot is paramagnetic whereas at high fields and temperatures the dot behaves
diamagnetic. An analogous situation is observed concerning the {\it
adiabatic temperature derivative} w.r.\ to changes of the magnetic field:
At low fields ($B<B_k$) and temperatures ($T<T_k$), the adiabatic temperature
derivative $\left (\frac{{\rm d}T}{{\rm d} B}\right )_S$ is $>0$, whereas it
is $<0$ at high fields and temperatures. For more results for GaAs
parameters, Fig.\ 2 and Fig.\
3 should be consulted, and it should be noted that in case of
Fig.\ 3 one obtains quite pronounced effects in the 'sensitive region' of
$T\approx 1.5$ K for $B\lsim 0.002$ Tesla.

\item In a many-electron planar dot ($N\gg 1$), the effects of quantization of orbital 
motion and the spin effects cannot be separated and should be treated
 simultaneously.  We do this within a simplified Hartree approach
 leading to a renormalized single-particle equation, where the effective
radius $R$ of the electron liquid on the dot is used as fitting parameter to
minimize the Free Enthalpy of the system at finite temperatures. In this
way, orbital and spin degrees of freedom are now coupled in a rather subtle
way.

 With increasing $B$, for $N\gg 1$, a spin-dependent
restructuring of energy levels takes place. As a consequence of a shell
effect, this leads to a {\it periodic} change of the
 magnetic
properties of the electrons in a dot with varying $B$, which is a function
of the variable $x:=\frac{\sqrt{N}\hbar|e|B}{2m^*\,c\,\eps_0}$ with period
$\Delta x=1$. (Here all parameters have their usual meaning, and
$\eps_0=\hbar \om_0$ is the confinement energy. It should be noted that the
period does not depend on the temperature and involves only the
characteristic energy scales of the system, except from the factor $\sqrt N$.
The factor $\sqrt{N}$ itself is of course related to the above-mentioned
condition of a half-filled outer shell, $N=m^2$.) Furthermore, the magnetic 
susceptibility of a dot with a half-filled last electron shell changes
 { not only by magnitude, but also by sign}: For low fields ($B<B_0$)
 and temperatures ($T<T_0$) the dot as a whole is paramagnetic, whereas
 with raising $B$ and $T$ the behaviour of the dot transfers from
 paramagnetic to diamagnetic. 

In an analogous manner, the temperature-effect induced by adiabatic
demagnetization of the dot behaves differently at low temperatures and
fields ($T<T_k$, $B<B_k$) and at high fields ($B>B_k$),
 respectively: In the first-mentioned case, the temperature diminishes
with an adiabatic decrease of $B$ whereas in the second case, it raises.
\ei
\subsection*{Acknowledgements}
This work was supported by project KBN PB 2 PO3B 055 18.
The first two authors would  like to thank the University
of Regensburg for hospitality.
\vglue 1 truecm
\centerline{\bf Appendix}

\noindent In this appendix we derive the relation between the periodic
function $ P_0(x)$, given by the 'Bloch representation' eq.\ (\ref{eq36}),
and the
corresponding 'Wannier representation' by the functions $A(x-n)$.

The periodic function $P_0(x)$ ($=P_1^{\,'}(x)$) is 

$$P_0(x)=2\sum\limits_{n=1}^\infty \tilde A(n)\cdot\cos(2\pi n
x)=\left \{\sum\limits_{n=-\infty}^{+\infty}\tilde A(n)e^{{\rm i}2\pi nx}\right \}
-1 \,,$$
where $$\tilde A(n)=\frac{2\pi^2nk_BT}{\eps_0}\left [{\rm sh}\left (
\frac{2\pi^2nkBT}{\eps_0}\right
) \right ]^{-1}\,.$$
Using the identity 
$$
\sum\limits_{n=-\infty}^{+\infty}\delta(x-n)=\sum\limits_{n=-\infty}^{+\infty}
\exp(-{\rm
i}2\pi n x)$$ we find
\beqa
P_0(x)&=&\left\{ 
\int\limits_{-\infty}^{+\infty}\tilde A(k)\cdot e^{2\pi{\rm
i}(x-n)}\cdot{\rm d}k
\cdot\sum_{n=-\infty}^{+\infty}\delta (k-n)\right \} -1\nonumber\cr
&=&\int\limits_{-\infty}^{+\infty}\tilde A(k)\cdot{\rm d}k\cdot
\sum\limits_{n=-\infty}^{+\infty}e^{2\pi{\rm i}(x-n)}
\,-1=\sum\limits_{n=-\infty}^{+\infty}A(x-n)\,-1\nonumber\,,
\eeqa
with
$$
A(x)=\int\limits_{-\infty}^{+\infty}\tilde A(k)\cdot e^{2\pi{\rm i}kx}
{\rm d}k\equiv\frac{\eps_0}{4k_BT}\left [{\rm ch}\left (
\frac{\eps_0x}{2k_BT}
\right ) \right ]^{-2}\,.
$$

\newpage

\centerline{\bf Figure Captions}
\vglue 1.0 truecm
Fig.\ 1: {The reduced magnetization $M(T,B)/\mu_B^*$ is presented as a
function of the temperature $T$ (in Kelvin units) and the magnetic induction
$B$ (in Tesla units) for a quantum dot with $N=1$ electrons on it, in a
two-dimensional electron gas with the parameters of GaAs, and with the confinement
potential parameter $\eps_0=\hbar\om_0=3.37$ meV. Note the change
of sign of $M$ from paramagnetic behaviour ($M > 0$) to diamagnetic
behaviour ($M<0$) by crossing the  contour line where
$M(T,B)\equiv 0$. (The apparent discontinuities of the contour lines,
representing 
$\frac{M(T,B)}{\mu^*}=-5\cdot 10^{-3}$, $-2.5\cdot 10^{-3}$,
$\pm 0$, $+2.5\cdot 10^{-3}$, ..., ,  as indicated at the margin,
 are due to inaccuracies of the plotting
software.)
}
\vglue 0.5 truecm

Fig.\ 2: {The 'adiabatic temperature derivative' $\left
(\frac{{\rm d}T}{{\rm d}B}\right )_S$ is presented against the temperature
$T$ (in Kelvins) and the magnetic induction $B$ (in Teslas) for a quantum
dot with $N=1$ electrons on it, in a two-dimensional electron gas with the
parameters of GaAs, and with the confinement potential parameter
$\eps_0=\hbar\om_0=3.37$ meV. Note the change of sign of the derivative from
positive values for low temperatures ($T<3$ K) to negative values for higher
temperatures, and note the strong increase in the region of 1.5 K for
inductions below 1 Tesla. (The apparent discontinuities of the contour lines
$\left (\frac{{\rm d}T}{{\rm d}B}\right )_S(T,B)=-1$, $-0.5$, $\pm 0$,
 $+0.5$,..., ,  as indicated at the margin,
  are due to inaccuracies of the plotting software.)}

\vglue 0.5 truecm

Fig.\ 3: {The same as in Fig.\ 2, but for inductions as low as 0.002 Tesla and below,
where the 'adiabatic temperature derivative' reaches extremely high values.
 (The apparent discontinuities of the contour lines
$\left (\frac{{\rm d}T}{{\rm d}B}\right )_S(T,B)=100$, $120$, $140$,
 ..., ,  as indicated at the margin,
 are due to inaccuracies of the plotting software.)
}
\newpage
\vglue 1.5 truecm
\epsfxsize = 12 cm
\epsfbox{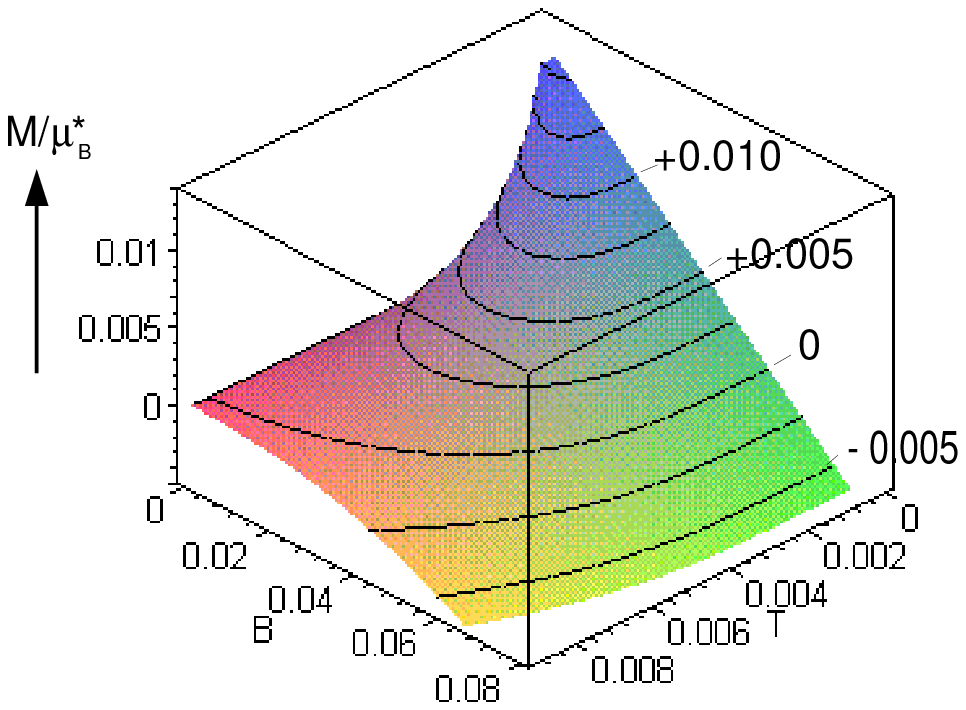}
\vglue 1 truecm

\centerline{Figure 1 :}
\vglue 0.3 truecm

\noindent
 {The reduced magnetization $M(T,B)/\mu_B^*$ is presented as a
function of the temperature $T$ (in Kelvin units) and the magnetic induction
$B$ (in Tesla units) for a quantum dot with $N=1$ electrons on it, in a
two-dimensional electron gas with the parameters of GaAs, and with the confinement
potential parameter $\eps_0=\hbar\om_0=3.37$ meV. Note the change
of sign of $M$ from paramagnetic behaviour ($M > 0$) to diamagnetic
behaviour ($M<0$) by crossing the  contour line, where
$M(T,B)\equiv 0$. (The apparent discontinuities of the contour lines,
representing $\frac{M(T,B)}{\mu^*}=-5\cdot 10^{-3}$, $-2.5\cdot 10^{-3}$,
$\pm 0$, $+2.5\cdot 10^{-3}$, ...,  as indicated at the margin,
 are due to inaccuracies of the plotting
 software.)

\newpage

\epsfxsize = 12 cm
\epsfbox{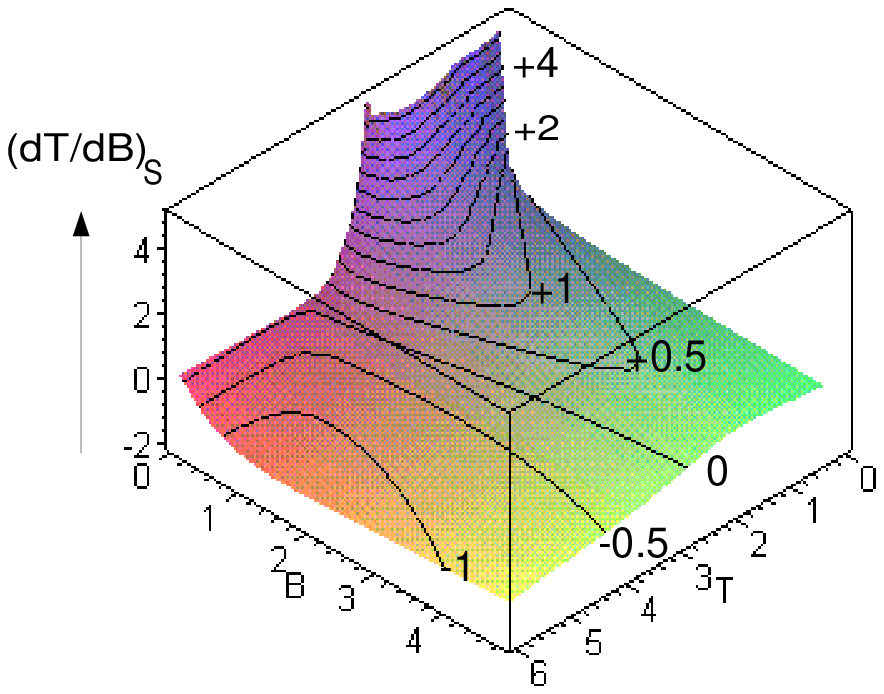}
\vglue 1 truecm

\centerline{Figure 2 :}
\vglue 0.3 truecm

\noindent
{The 'adiabatic temperature derivative' $\left
(\frac{{\rm d}T}{{\rm d}B}\right )_S$ is presented against the temperature
$T$ (in Kelvins) and the magnetic induction $B$ (in Teslas) for a quantum
dot with $N=1$ electrons on it, in a two-dimensional electron gas with the
parameters of GaAs, and with the confinement potential parameter
$\eps_0=\hbar\om_0=3.37$ meV. Note the change of sign of the derivative from
positive values for low temperatures ($T<3$ K) to negative values for higher
temperatures, and note the strong increase in the region of 1.5 K for
inductions below 1 Tesla. (The apparent discontinuities of the contour lines
$\left (\frac{{\rm d}T}{{\rm d}B}\right )_S(T,B)=-1$, $-0.5$, $\pm 0$,
$+0.5$, ..., ,  as indicated at the margin,
are due to inaccuracies of the plotting software.)}
\vglue 0.5 truecm

\newpage

\epsfxsize = 14 cm
\epsfbox{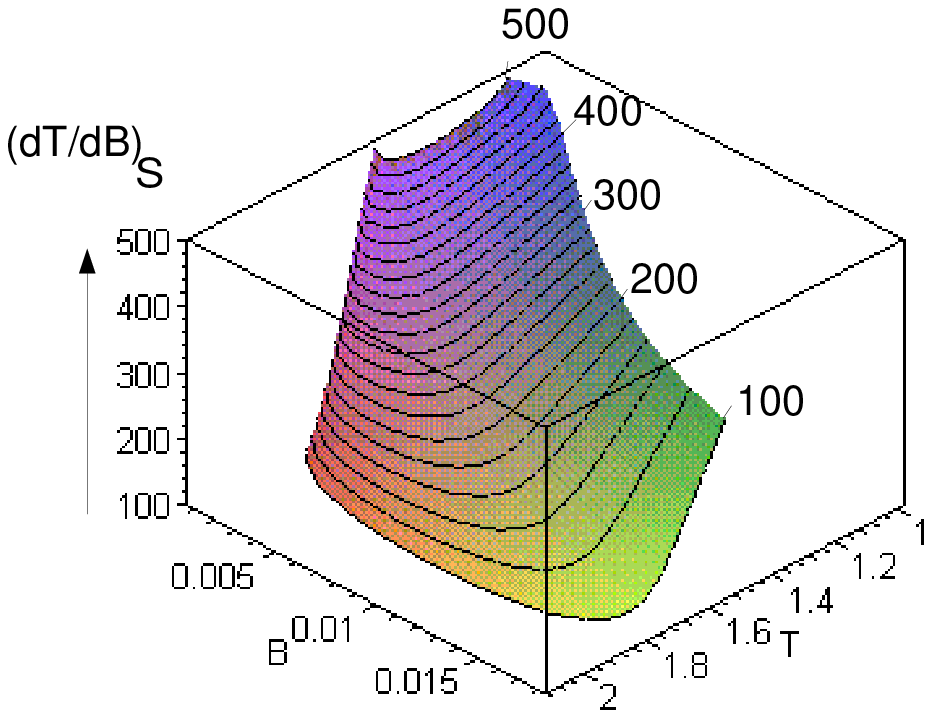}
\vglue 1 truecm

\centerline{Figure 3 :}
\vglue 0.3 truecm

\noindent
{The same as in Fig.\ 2, but for inductions as low as 0.002 Tesla and below,
where the 'adiabatic temperature derivative' reaches extremely high values.
 (The apparent discontinuities of the contour lines
$\left (\frac{{\rm d}T}{{\rm d}B}\right )_S(T,B)=100$, $120$, $140$,
 ..., ,  as indicated at the margin,
  are due to inaccuracies of the plotting software.)

\begin{thebibliography}{99}
\bibitem{refREM1} Since the Zeeman term '$-B\cdot M$' is included in the Hamiltonian $\cal H$,
the expectation value $\langle{\cal H}\rangle$ is called 'Internal Enthalpy'
whereas the name 'Internal Energy' is reserved for the expectation value of
those parts of $\cal H$, e.g.\ the 'exchange energy',
 which do depend on $M$, but not explicitly on $B$.
In fact, the relation between the 'Internal Enthalpy' $I(T,B)$  and the
'Internal Energy' $U(T,M)$ is given by the {\it Legendre transform}
 $I(T,B)=U(T,M)-B\cdot M$.
\end{thebibliography}
\end{document}